\documentclass[twocolumn,showpacs,aps,amsmath,amssymb,prl,superscriptaddress,nofootinbib]{revtex4-1}

\usepackage{graphicx}
\usepackage{amssymb}
\usepackage{amsmath}
\usepackage{comment}
\usepackage{epstopdf}
\usepackage{epigraph}
\usepackage{varioref}
\usepackage{hyperref}
\usepackage{cleveref}

\usepackage{comment}
\usepackage{color}
\newcommand{\bea}{\begin{eqnarray}}
\newcommand{\eea}{\end{eqnarray}}

\usepackage{color}
\begin{document}
\title{Artificial Spin Ice Phase-Change Memory Resistors}
\author{Francesco Caravelli}
\email{caravelli@lanl.gov}
\affiliation{Theoretical Division and Center for Nonlinear Studies,\\
Los Alamos National Laboratory, Los Alamos, New Mexico 87545, USA\\
}

\author{Gia-Wei Chern}
\email{gc6u@virginia.edu}
\affiliation{Department of Physics, University of Virginia, 
Charlottesville, VA 22904, USA}

\author{Cristiano Nisoli}
%\email{cristiano@lanl.gov}
\email{cristiano@lanl.gov}
\affiliation{Theoretical Division and Center for Nonlinear Studies,\\
Los Alamos National Laboratory, Los Alamos, New Mexico 87545, USA\\
}

\begin{abstract}
We study the implications of the anisotropic magnetic resistance on permalloy nanowires, and in particular on the property of the resistance depending on the type of lattice. 
We discuss how the internal spin configuration of artificial spin ice nanowires can affect their effective resistive state, and which mechanisms can introduce a current-dependent effect dynamic resistive state. We discuss a spin-induced thermal phase-change mechanism, and an athermal domain-wall spin inversion. In both cases we observe memory behavior reminiscent of a memristor, with an I-V hysteretic pinched behavior.
\end{abstract}

\maketitle
%\section{}
%\subsection{}
\textit{Introduction.} The study of interacting magnetic nanostroctures called {\it artificial spin ices} \cite{colloq,wang0,Bader,reddim,Heyderman,Canals1,Chioar1,Nisoli1,Nisoli2} has now reached a level of control~\cite{Nisoli4,Gilbert2,Bhat,Tierno1,Tierno2,Tierno3,Latimer,gartside,vavassori,WangYL3} that should open the way to technological applications. To this day they have been employed to study geometric frustration, ground state degeneracy, dimer excitations, and a tendency towards topological order~\cite{Castelnovo1,Mengotti,topor,Chern2,Gliga} and generally as a setting to generate exotic states and behaviors. 

Since spin ice materials encode naturally internal states in some observable systemic phenomena, it has been suggested that these meta-materials can be engineered for the purpose of logical computation~\cite{GilbertMem,Lammert2,CaravelliNisoli,Arava,CASI}. Yet, another interesting venue in computation is to use artificial spin ice to design resistive memory and possibly memristors. The use of memory effects in resistive materials have been suggested for a variety of computing applications \cite{spin1,spin2,spin4,spin5} and resistive switching \cite{memr1,memr2,memr3,traversa,memrc1,memrc2,inm1,inm2} (see  \cite{ProbComp5} for a broad introduction, or \cite{review1,review2} for a more technical one). Boolean logic computation has also been proposed via experiment proof of principle logic gates \cite{bernstein1,bernstein2,bernstein3,Arava,CASI,bernstein4} or via hierarchical gate integration proposals \cite{CaravelliNisoli}.\\

Here we explore the possibility of using artificial spin ice to engineer memristors, based on previous work on transport and magnetoreristance in these materials~\cite{amrsi,amr,chir}, and start from  the theoretical framework previously developed by one of us~\cite{gwc}. It was shown there that connected artificial spin ice is as a electrical circuit where tension drops at the vertices because of the magnetoresistive effects of domain walls there. Vertices can be considered electrical elements whose functionality is controlled by the magnetic moments impinging in them: changing their configuration affects the resistance of the system, leading to reconfigurable circuitry. However one can imagine that current itself could alter the moment configurations, thus leaving memory of its passage. We consider two possibilities for this coupling. One is based on Joule effect, where the superparamagnetic temperature of the nanoislands exceeds by little the operative temperature of the system~\cite{thermal1}. The other is through the spin torque of a spin polarized current~\cite{spininversion1,spininversion2,spininversion3,spininversion4}.

\begin{figure}[b!]
 \includegraphics[scale=1.5]{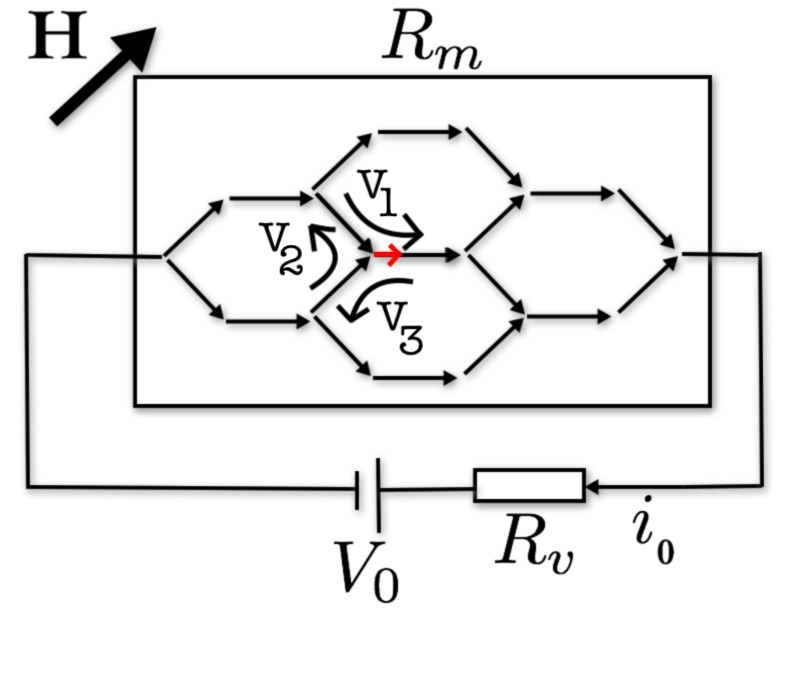}
\caption{ The structure of the anisotropic magneto-resistance memristor device we study, within a Kagome lattice. We consider a system of resistive nanowires of magnetic permalloy, which are connected to a battery. The currents flow in the nanowire, but because of the anisotropic magnetoresistance induced by a small external magnetic field $H$, the internal resistivity depends on the distribution of the magnetic moments in the wires at the junctions. 
}
\label{fig:conf}
\end{figure}

In this work we explore whether connected artificial spin ice can function as a memristor by solving the collective dynamics of currents that alter the magnetic texture, which in turns alters the localized resistance and then the currents themselves.  We first derive a perturbative equation for the effective resistance of the device as a function of the nanoislands moment configurations. We then use this exact solution to obtain via self-consistency a closed equation for the conductance of the device. This latter can be written as the sum of the conductance of the permalloy nanowires and a state dependent conductance. We show first the simple case of a 3-legs junction, and then show that in general for materials with a sharp magnetic order transition the resistance is of the phase-change type. We then extend our study by simulating a Kagome lattice \cite{honey1,honey2} when a threshold domain wall spin inversion is considered, and observe the effect of the many-body interaction on the resistance. 

%When a small magnetic field is applied to the permalloy two different resistances develop along the directions parallel ($\rho_{||}$) and the orthogonal ($\rho_{\perp}$)  to the magnetic field. For the case of a permalloy, a small magnetic field of a few Oersted suffices \cite{amrsi}. %We are in particular interested in the collective galvanomagnetic phenomena that are due to the ordering of the magnetic spin of the islands when the system is below the Curie temperature.\\
Consider a network of permalloy nanowires, as shown in Fig. \ref{fig:conf}. Each wire portion between vertices is magnetized and the coupling between moments is such to obey the ice-rule.  If the external magnetic field is zero, each  wire of the network will have a resistance $R=\rho_0 L$, where $L$ is the length of the island. %We assume for simplicity that the resistances are identical across the circuit.
The presence of magnetization $\vec m $ alters the resistance according to the Anisotropic MagnetoResistance (AMR) law at low magnetic field, $H$
\begin{equation}
\vec E=\rho_0 \vec J+\hat m ( \rho_{||} -\rho_{\perp}) (\hat m\cdot \vec J)
\end{equation}
where $\vec J$ is the density of current, and  $\rho_{||}(H)$ and $\rho_{\perp}(H)$ are the resistances parallel and perpendicular to the magnetization, respectively (and $\Delta \rho=\rho_{||} -\rho_{\perp}$).
Along a line $\gamma$ in the material, the voltage drop is 
\begin{equation}
V_{\gamma}=\int_{\gamma} \vec E\cdot d\vec t  %\nonumber \\
%=\int_{\gamma} \left(  \rho_0 \vec J+\hat m \Delta \rho (\hat m\cdot \vec J)\right)\cdot d\vec t\nonumber \\
= V_0+\Delta \rho \int_{\gamma}   \left(\hat m\cdot \vec J\right) \hat m\cdot d\vec t,
\label{eq:amr}
\end{equation}
and as  will shall see it affects the behavior of the device by introducing a state-dependent memory. 
As noted previously by two of us~\cite{gwc,amrsi}, the AMR is independent from a global  change of spin configuration. % $\vec m\rightarrow -\vec m$, and this implies invariance of the voltage configurations with respect to $\vec s\rightarrow -\vec s$, while the effect only comes from the vertices. 

If $V_0$ is an external potential applied to the circuit in one direction (see Fig.~1), the current follows Ohm's law, $i={V_0}/(R_v+R_m)$, where $R_m$ is the resistance of the galvanomagnetic material of interest and $R_v$ is an external resistence.  We aim to show that the internal configuration of the magnetization acts as a memory for the resistance $R_m$. 
We neglect any parasitic capacitance. %The behavior of the Kirchhoff's voltage law is however modified via the extra term in eqn. (\ref{eq:amr}) and due to the anisotropic  magnetoresistance effect. 
We assume that the magnetization of the system is at a equilibrium temperature at a certain temperature $T$.

Two of us showed previously~\cite{gwc,amrsi} that the voltage drop across nodes depends on the configuration of the moments in the nanowires $\{s_i\}$. The problem then becomes how to obtain the current distribution when the magnetic moments configuration is known. For that we employ a graph theoretical approach~\cite{bollobas,memrc1}, which has been already successful in the study of circuits of memristors \cite{memrc1,memrc2,Caravelli2,Caravelli3}. 

Consider a general graph $G$ (for definitness a Kagome spin ice in Fig.~1) with $N_v$ vertices (or nodes) and $N_e$ edges, which describes a network of resistors. The graph supports $N_c$ loops (closed loops or subcircuits). One can describe the potential equivalently by assigning a potential $p_\alpha$ at each node, or a potential drop $v_k$ for each edge $k$ hosting a current $i_k$  (we use latin indices for the edges, and greek indices for the nodes, greek indices with tildes represent instead cycles on the graph). We choose an orientation for the current on each edge--something that can be done in $2^{N_e}$ ways--and encode it into the matrix $B_{\alpha k}$ of size $N_v \times N_e$. 

In this language, $\sum_{j=1}^M B_{\alpha j} i_j=(B \cdot \vec i)_{\alpha}=0$ enforces  Kirchhoff current law at each vertex $\alpha$. Then the potential drop $v_k$ for each edge $k$ along the chosen direction is given by $v_k= \sum_{\xi} p_{\xi} B_{ \xi k}=(^t\vec p \cdot B)_k $. 

The Kirchhoff Voltage Law (KVL) can be written as $\sum_k A_{\tilde \xi k} v_k=0$ where $A_{\tilde \xi m}$ is the $N_c \times N_e$ cycle or loop matrix, obtained by assigning first an orientation to the edges and then the loops of the graphs, and assigning values of $+1,0,-1$ if the orientation of the loop agrees or disagrees with the orientation of the edge. This equation simply states that the circuitation of the voltage on voltage on a node must be zero. As a consequence, in general, $B\cdot ^tA=A\cdot ^tB\equiv 0$. Finally we introduce $R$ the $N_e \times N_e$ diagonal matrix of resistances. 

As shown in ref.~\cite{memrc1,bollobas} one obtain a generalization of the Ohm's law in the form
\begin{equation}
\vec i=- A^t (A R A^t )^{-1} A \vec V(\{s\})=Q \vec V(\{s\}).
\label{eq:sol0}
\end{equation}
where we defined the symmetric matrix $Q=-A^t (A R A^t )^{-1} A$.
The vector $\vec V$ is the vector of effective voltages in series to the resistances on the graph (see the Supplementary Material, SM-C, for details).
Its evaluation can be carried out analytically given a certain lattice configuration. In the SM-D, an analytical expression in implicit form is provided for the Kagome lattice, which will be used shortly. 

But first we discuss the implications of eqn. (\ref{eq:sol0}) when an external voltage is applied to the material, as in Fig. \ref{fig:conf}. 
The internal currents depend on the voltage, which in turn depends on the spin configuration within the material, and linearly in the currents as from eqn. (\ref{eq:amr}). Then, a self-consistent nonlinear equation for the voltages can be obtained. Since the anisotropic magnetoresistance is a small effect (it contributes ~3-5\% on the material resistance) we can linearize Eq.~\ref{eq:sol0} in  $\Delta \rho$. Remarkably, the final equation for the effective resistance is simply a parallel resistor equation (see SM-C and SM-D):
\begin{equation}
  R^{-1}\equiv G=G_0+G_{m}(s),
\end{equation}
where $G_0$ is the conductance when no anisotropic magnetic resistance is present, and $G_m(s)$ is a state-dependent function with the dimension of an inverse resistance. In particular, $G_{m}(s)=\Delta \rho \vec Q^t M(s) \vec Q$ is a quadratic form where $\vec Q$ is a network dependent vector with the dimension of inverse resistance and can be obtained from the matrix $Q$ from the rows (or column) corresponding to the resistance in parallel to the generator and the internal resistances. Instead, $M(s)$ is an adimensional matrix which depends on the internal magnetization state via terms of the form $s_i s_j$. It is hard to obtain exact expressions for $\vec Q$, but this can be easily obtained numerically. %The exact derivation of $G_m(s)$ is provided in a worked out example in SM-D.
Naturally $M(s)$ changes in time with the currents, leading to a memory effect that we aim to show to be memristive. How it changes depends on an underlying physical mechanism. 

%One simple way to visualize why the spin configuration changes the effective resistance, let us consider a toy model.
%We consider two aligned spins on a single mesh, with a certain external voltage $V_0$. The effective voltage depends on the state of the two spins and proportionally to the current, as noted in eqn. (\ref{eq:amrcurr}). We assume that the effective voltage in the wire depends on the whether the two spins are aligned or not, and thus effectively on the quantity $s=s_1+s_2$. Ohm's law implies
%\begin{equation}
%    $V_0+V_{e}(s)\equiv V_0+ r_0(s) i = R i$. 
%\end{equation}
%We can thus $V_0= (R-r_0\left(s(t)\right) ) i$ and $r_0$ is the spin-state dependent effective resistance.  If the effective resistive state depends on the state of the spin, if we introduce a current-dependent spin flipping mechanism, we can rewrite the resistance in terms of two limiting resistances $R_1$ and $R_2$,
%\begin{equation}
%    $r_0(t)=R_1 + R_2 \theta\left(I(t)-I_c\right)$,
%\end{equation}
%which is effectively a switch. The toy model above simply explains why the system is memristive.

%

%

\begin{figure}
    \centering
    \includegraphics[scale=0.3]{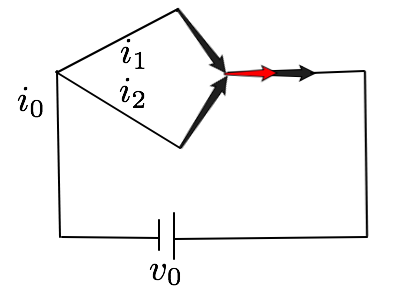}
    \caption{A simple example of the internal memory of the magnetoresistance.}
    \label{fig:simple}
\end{figure}

We posit that if the nanoislands are close enough to the superparamagnetic threshold they can become thermally active due to teh Joule effect of the applied current. %We initially performed a Glauber dynamics simulation for a small system with a simple 3-spin Hamiltonian in order to show that the thermal coupling implies resistance memory (and provided in the SM-E). The temperature is introduced via Joule heating and a regulation mechanism like Stefan's radiation.  However, we can introduce an effective model which depends only on an equilibrium thermal correlation functions $\langle s_i s_j\rangle_T$ obtained from a many-body or spin-current interaction.
For illustration we consider a simple 3-moments system, as in Fig. \ref{fig:simple}. In this case, the directionality of the moment now take a rather small set of possible resistances which introduces different resistive states. For larger system however we can introduce an effective description for the resistance.

\begin{figure}
    \centering
    \includegraphics[scale=0.37]{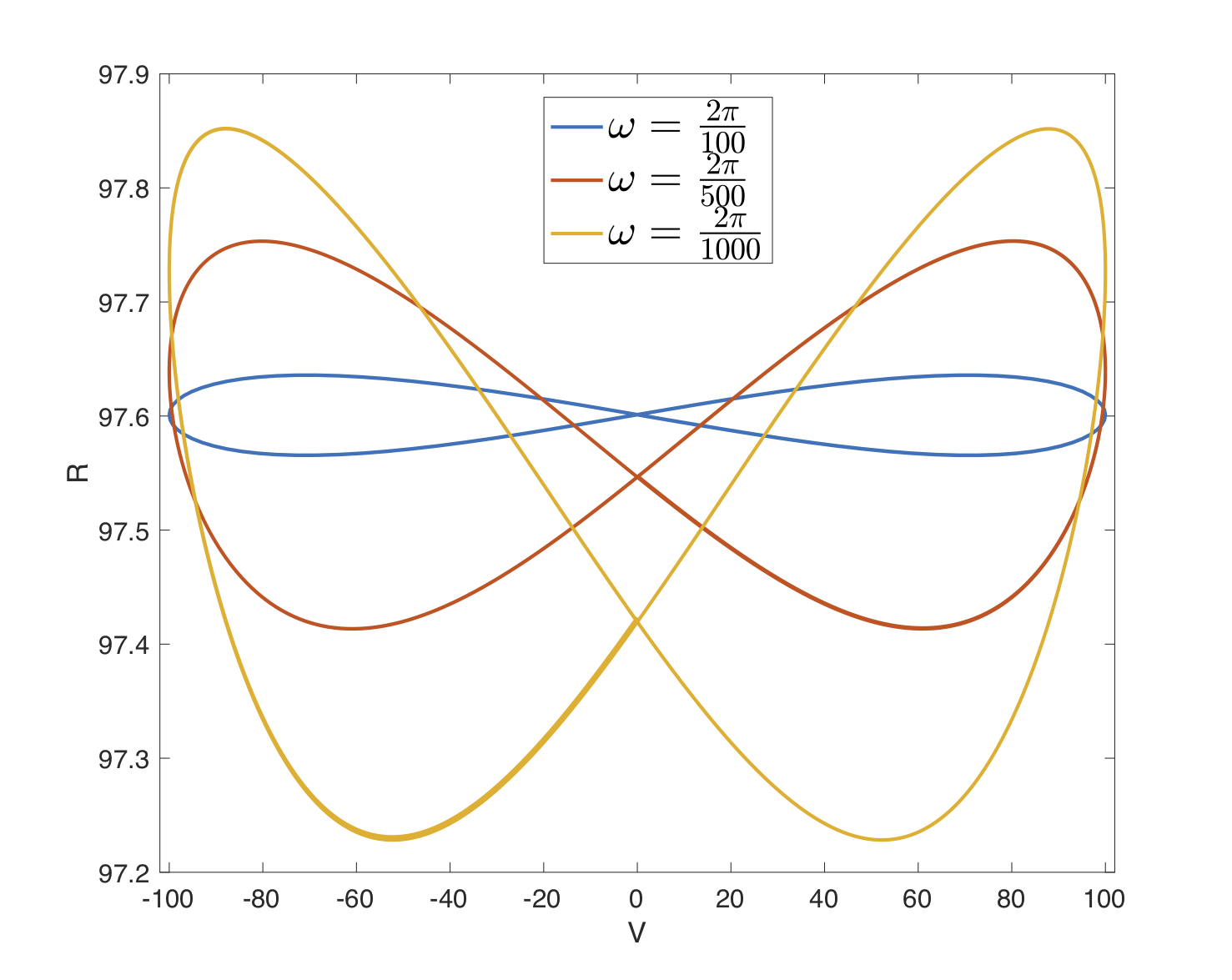}
    \caption{Memristive behavior shown on the dependence of the Lissajous figure for the resistance as a function of the voltage, on the frequency of the sinusoidal input $V=V_0 \sin(\omega t)$ for a ``crossover" transition, with a smoothed $\theta_k(x)$ function for $k=1$. We plot in particular $R/R_0$ but for $R_0=1$ and in units in which $V_0=1$. %\textcolor{red}{Magari nella caption rimpiazza $V$ con $V/V_0$ e $R$ con $R/R_0$}
    }
    \label{fig:lissajous}
\end{figure}

We can write an approximate (first order contribution) equation for the thermal average of the effective resistance of the form:
\begin{equation}
    \langle R^{-1}\rangle_T=R_0^{-1}+\Delta \rho \left( \theta_k(T-T_c)R_{<}^{-2}+\theta_k(T_c-T)R_{>}^{-2}   \right)
\end{equation}
where $R_{<}$ and $R_{>}$ are two resistances which depend on the value of $M(s)$ and the geometry of the spin ice (which is contained in $\vec Q$).  The resistances are defined as quadratic forms (see SM-C)
\begin{eqnarray}
    R_{<}=\vec Q\cdot M_{<}\vec Q, \ \ \ \ \
    R_{>}=\vec Q\cdot M_{>}\vec Q.
\end{eqnarray}
where $M_{<}\equiv M\Big(\langle s_i s_j\rangle=0\Big)$ and $M_{>}\equiv M\Big(\langle s_i s_j\rangle=1\Big)$, and $\vec Q$ is a circuit dependent vector.  The two limiting resistances can be obtained also if the interaction is antiferromagnetic (and the graph bipartite). The smoothed theta function, in this case, is a reasonable smoothed $\theta_k$-function is $\theta_k(x)=\frac{1}{1+e^{-k x}}$, where $\theta(x)=\lim_{k\rightarrow \infty} \theta_k(x)$ which incorporetates a sharp transition or a crossover. The effective mechanism is based on the fact that that the equilibrium temperature depends on the balance between radiation and current induced Joule heating.
Thus, in the simplest possible approximation, we see that under the application of a small magnetic field the resistance of the material can be assumed to be a thermal phase-change type of material (or switches), in which the system has two resistance phases depending on the current, which in turn controls the temperature of the permalloy via Joule heating. If the transition is not sharp but only a crossover, then we can assume that
\begin{equation}
    \langle R^{-1}\rangle_T=R_0^{-1}+\frac{\Delta \rho}{  \tilde R(T)^{2}}   
\end{equation}
where $\tilde R$ is a smooth function such that $\tilde R \to R_{<}$ as $T\to o$ and  and $\tilde R \to R_{>}$ as $T\to \infty$. In this case, it seems reasonable to assume that the material will fall in the thermal memristor framework introduced in \cite{thermal1}.
Then, because we expect the typical memristive $v-i$ hysteresis to be small, it is possible to see the change in the resistances from the $v-r$ Lissajous figures (Fig.~2), obtained from the functional dependence of the effective resistance which we have obtained.

\begin{figure}
    \centering
    \includegraphics[scale=0.17]{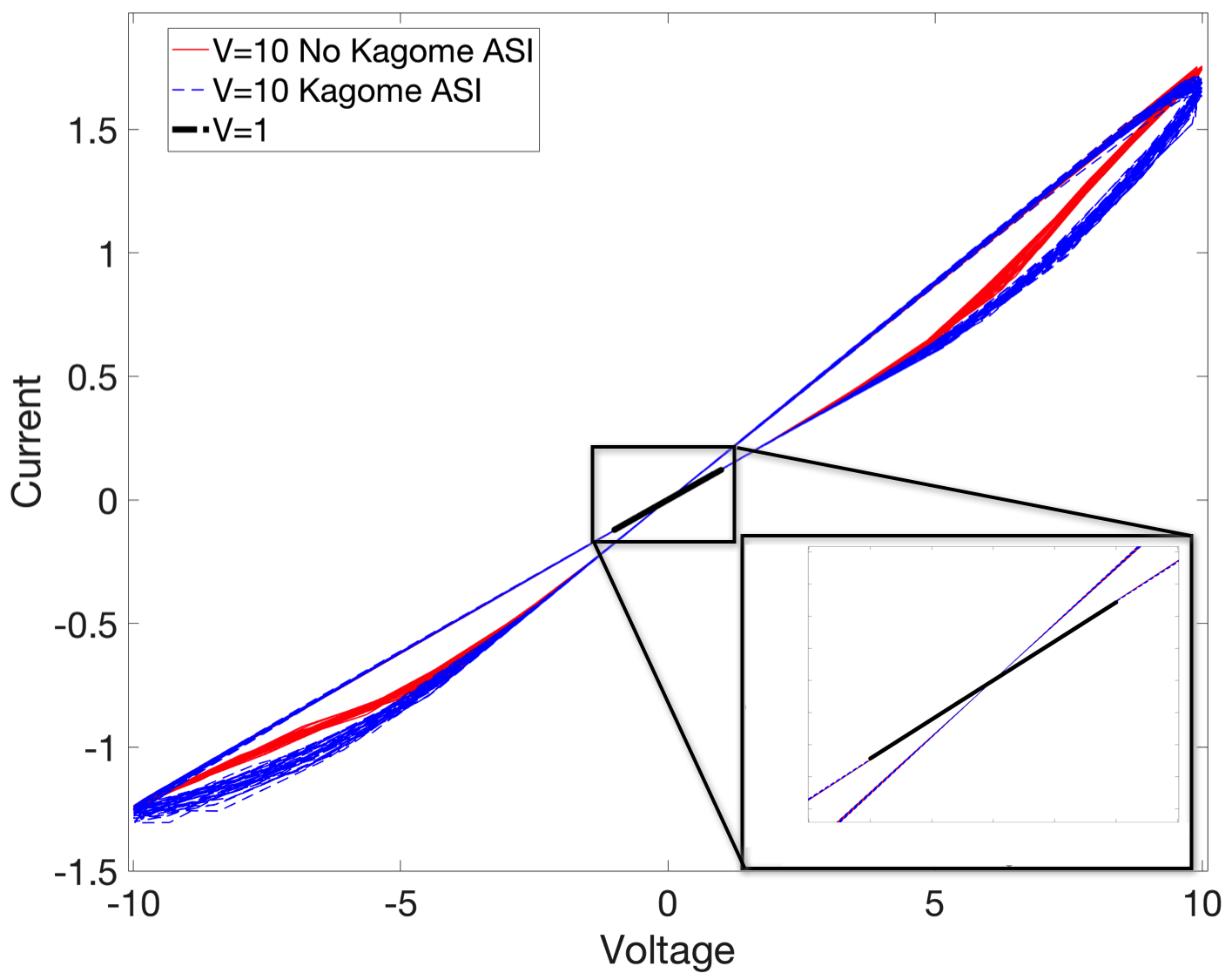}
    \caption{Memristive behavior of honeycomb spin ice from  the spin-dynamics simulations. Current vs. voltage curves for the interacting (blue) and non interacting (red) system for $R_0=1$,$\Delta\rho=0.1$, $I_{c}=0.1$ $M=L=8$, $T=1000$, $dt=0.1$, $\omega=30$ for $V_0=10$ and $V_0=1$. The initial condition is a honeycomb ice in the completely ordered state. In inset, result for the non-interacting system at small $V_0=1$ shows lack of hysteresis.%\textcolor{red}{Vedi la precedente per le labels. togli la legenda interna. è difficile leggere i numeri nell'inset. Perchè hai scelto $\Delta \rho=0.1$ quando può essere tranquillamente 3? La curva di isteresi sarebbe stata più ampia}
    }
    \label{fig:hyst}
\end{figure}

%Albeit the exact numbers will depend on the type of material, we expect to be able to distinguish the type of transition from the $V-R$ curves of the device as a function of the frequencies. At slower frequencies, the changes in the resistance will be more symmetric in the continuous case, while more abrupt but still hysterestic in the discontinous case. The mechanism above is a well known phase-change type of material mediated by spin-spin interactions. However, we argue below that in permalloy nanowires there can be memory also via athermal mechanisms. \textcolor{red}{Questo paragrafo non mi è molto chiaro. Che tipo di transizione? In che modo questo si riferisce alla Fig 2? }

Another mechanism for moment inversion in nanowires which does not require  very careful fine tuning of the temperature, is the  
 current-induced domain wall inversion via spin-transfer~\cite{spininversion1}.  Specifically, when a current is applied to a magnetic nanowire, some domain wall defects can form at the junctions and quickly move along the wires \cite{spininversion2,spininversion3,spininversion4}. This phenomenon can be effectively modeled in our system by assuming that if the current in a wire is higher than a certain threshold $I_c$, then the spin along that wire is inverted. 

In a large system this hard switching behavior gets smoothed out. We consider an extended Honeycomb circuit of 16 loops, where heach whire has resistence $R_0$. We then consider the following two-steps spin dynamics. At each time step we start with a spin configuration $\vec s_{t-1}$, solve for the Kirchhoff laws in the nanowires, and find the equilibrium currents $\vec i(t)$ in the material as a function of the external voltage $V(t)= V_0 \cos(\omega t)$ only. We then use the eqns derived in the SM-C to find the auxiliary voltage generators (given the spin configuration $\vec s_t$) in the material which will affect the spin configuration $\vec i(t)$. At this point, we consider the domain-wall inversion process. If the current in each wire is above a certain threshold $I_c$ and the current is in the opposite direction of the magnetization direction of that wire, we flip the magnetization direction instantaneously. In experiments like those of \cite{spininversion1} the switching is extremely fast. We thus consider for simplicity instantaneous inversion.

We account for the manybody interaction among spins by imposing constraints on the possible vertex configurations. Honeycomb spin ice \cite{gwc,honey1,honey2,gwc2} is frustrated and at low energy enters an ice-rule regime where only vertices with two moments pointing in and one out, or viceversa, are allowed. Therefore we only allow spin inversion when it produces vertices of  magnetic charge equal to $Q=\pm 1$, and neglect the nodes with $Q=\pm 3$.

Once we fixed all the parameters, there will be a threshold $V_c$ which depends on the size of the system and the resistivity of the material above which the spin-inversion occurs and below which the system is a normal resistor. The threshold dynamics is reminishent of a fuse-network dynamics \cite{Forrest}, but with the difference that it is not the conductance that dramatically drops, but that the effective voltages change instead.

In Fig. \ref{fig:hyst} we plot resultsof current vs. voltage. For small values of $V_0$ no memristive behavior is present (inset), as expected.  For both the non-interacting (red) and interacting (blue) honeycomb lattice we obtain a  zero-crossing pinched hysteresis loop typical of memristive devices and which suggests the presence of memory  \cite{inm2,ProbComp5,review1}. The latter is more hysteretic and smoother than the former but memristive behavior is already present even in absence of interactions. The area of hysteresis is small due to the small value of the magneto resistence effect.

We have put forward a theoretical framework for the study of memory properties of magnetic nanowires subject to an external field as an effect of anisotropic magnetoresistance, building on previous results \cite{gwc}. We have derived exact and general equations which show, given a certain spin ice lattice, how the resistance of the material changes given the internal configuration. This has enabled us to obtain first order contributions to the resistance, showing that there exist an effective resistance in parallel to the nanowires network and which depends on the internal state of magnetization. Then teh coupling between current and magnetism lead to a memristive behavior. We studied two mechanisms which induce these memory effects, and which likely coexist. As the anisotropic magnetoresistance effect is of the order of a few percentages, we expect a smaller change in the value of the resistance as a function of the voltage. However, we have shown that the hysteresis curves depend on the many-body interaction and the configurational manifold of artificial spin ices. This suggests that more generally the functionality of artificial spin ice memristors can be open to design--as so many other properties of these materials proved to me. These ideas are an alternative to Spin-Torque memristors for bio-inspired computing \cite{grollier}, to produce an effective magnetic phase-change material \cite{pcm}. In fact, because of the sensitivity to temperature, memory resistors in spin ices can have a variety of behaviors that can serve as an alternative to known Spin-Transfer-Torque devices \cite{grollier,stiles} and Phase-Change Materials \cite{pcm}. Furthermore, as the magnetic moments can be acted upon collectively by external magnetic field, or individually, artificial spin ice memristors could be reprogrammable. \\

 The work of FC and CN was carried out  under the auspices of the US Department of Energy through the Los Alamos National Laboratory, operated by Triad National Security, LLC (Contract No. 892333218NCA000001). CN was founded by DOE-LDRD grant 2017014ER. FC was also financed via DOE-LDRD grants PRD20170660 and PRD20190195.

\end{document}

\title{Artificial Spin Ice Phase-Change Memory Resistors: Supplementary Material}
\author{Francesco Caravelli}
\email{caravelli@lanl.gov}
\affiliation{Theoretical Division and Center for Nonlinear Studies,\\
Los Alamos National Laboratory, Los Alamos, New Mexico 87545, USA\\
}

\author{Gia-Wei Chern}
\email{gc6u@virginia.edu}
\affiliation{Department of Physics, University of Virginia, 
Charlottesville, VA 22904, USA}

\author{Cristiano Nisoli}
%\email{cristiano@lanl.gov}
\email{cristiano@lanl.gov}
\affiliation{Theoretical Division and Center for Nonlinear Studies,\\
Los Alamos National Laboratory, Los Alamos, New Mexico 87545, USA\\
}
\maketitle
\section{Supplementary MAterial}
\subsection{A. Formal solution of linear circuits} \label{sec:derivationstart}
We use a graph theoretical approach~\cite{bollobas,memrc1} to solve for the current knowing the nanoisland moments. Consider a graph $G$ (for definitness a Kagome spin ice in the figure) with $N_v$ vertices (or nodes) and $N_e$ edges, which describes a network of resistors. The graph supports $N_c$ cycles, that is closed loops or subcircuits.  In each node there is a potential $p_\alpha$, and for each edge a current $i_k$  (we use latin indices for the edges, and greek indices for the nodes, greek indices with tildes represent instead cycles on the graph). We choose an orientation for the current on each edge--something that can be done in $2^{N_e}$ ways and encod it into the matrix $B_{\alpha k}$ of size $N \times M$. Then $\sum_{j=1}^M B_{\alpha j} i_j=(B \cdot \vec i)_{\alpha}=0$ enforces Kirchhoff current law at each vertex $\alpha$. Then the potential drop $v_k$ for each edge $k$ along the chosen direction is given by $v_k= \sum_{\xi} p_{\xi} B_{ \xi k}^t=(^t\vec p \cdot B)_k $.

The Kirchhoff Voltage Law (KVL) be written as $\sum_k A_{\tilde \xi k} v_k=0$ where $A_{\tilde \xi m}$ is the $N_c \times N_e$ cycle or loop matrix. This equation states that the circuitation of the voltage on voltage on a node must be zero. As a consequence, in general, $B\cdot ^tA=A\cdot ^tB\equiv 0$. 
In order to see how formally one can introduce the reduced loop matrix, such that $(ARA^t)$ is invertible, we need some notion of graph theory. Given the graph $G$, we introduce a spanning tree $\mathcal T$ (called \textit{co-chords}), and the set of edges of the graph not included in the tree as $\mathcal T$, \textit{ or chords}, are given by $\bar {\mathcal T}$. For each element of the chord $\bar {\mathcal T}$, we assign a cycle, called \textit{fundamental loop}.
 The loop matrix $A$, can be reduced to its $N_e-N_v+1$ {\it fundamental}  loops. Then, it is not hard to show that the current vector can be written as $\vec i=(A_{\mathcal T}^t {\vec i}_c, {\vec i}_c)=A^t {\vec i}_c$,
 where we used the fact that given a chords and co-chords splitting, we have $(B_{\mathcal T}, B_{c})\cdot (A_{\mathcal T},\ I)^t=0$, which implies $A^t_{\mathcal T}=- B_{\mathcal T}^{-1} B_{c}$.  Since $A$ is derived from the \textit{reduced} incidence matrix, this is called \textit{reduced loop matrix}.
At this point, it can be shown that 
%If we now write the equation for the circuit, i.e., $\vec v=R \vec i+ \vec S(t)$, we note that applying the operator $A$ on the left, we obtain the identity $A \vec v=0 =A R \vec i+A \vec S(t)$. We now use ${\vec i}=A^t {\vec i}_c$, and obtain $(A R A^t) {\vec i}_c=- A \vec S_{0}(t)$. If we now write the solution of the current, we obtain 
\begin{equation}
\vec i=A^t {\vec i}_c=- A^t(A R A^t)^{-1} A \vec S(t). 
\label{eq:init0}
\end{equation}
which is the starting point of the paper. It is not hard to see that $AR A^t$ is always invertible for non-zero resistances. For more details, we refer to \cite{bollobas,memrc1}.

The reduced loop matrix $A$ is constructed using the following procedure. First, we assign an orientation to the edges of the graphs, and for each loop of the graph, we assign an arbitrary orientation to the loop along each edge of the loop. We then construct the matrix $A_{L E}$ (dimensions of loops by edges) as follow. If the loop $N_c$ does not contain the edge $E$, $A_{LE}=0$. If the orientation of the loop agrees with the orientation of the edge, then $A_{LE}=1$, otherwise $A_{LE}=-1$. At this point, we choose a subset of $N_e-N_v+1$ linearly independent loops and remove the others from $A$. What we obtain is the reduced loop matrix.

\subsection{B. Mapping voltages drops at nodes to voltage generators}
It is common in spin ice materials to approximate the magnetization with a internal configuration $\mathcal M=\{\vec s_i\}$, where $ \vec s_i =s_i \{a \hat x, b \hat y \}$ are Ising variables on the plane, which cannot rotate.
\begin{figure}
\includegraphics[scale=1]{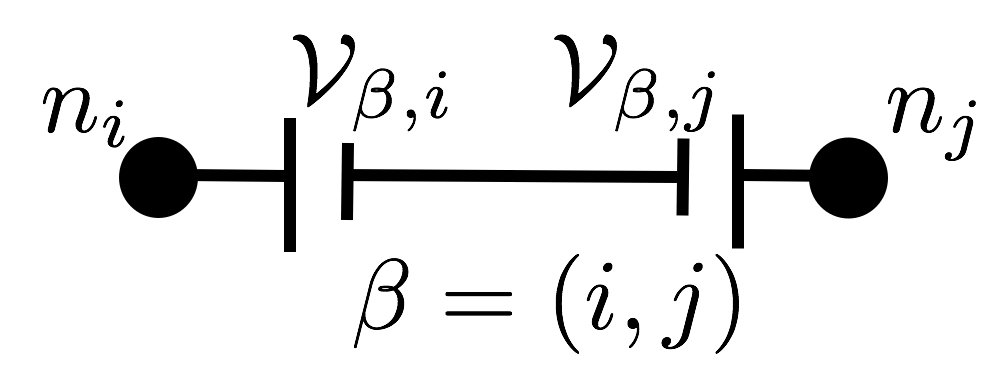}
\caption{The edge voltage configuration: for each node, there is an associated voltage.}
\label{fig:edgeconf}
\end{figure}
The plan is to map the node configuration to a set of voltages in series to the resistances, as this is an exactly solvable model
\begin{figure}
\includegraphics[scale=1]{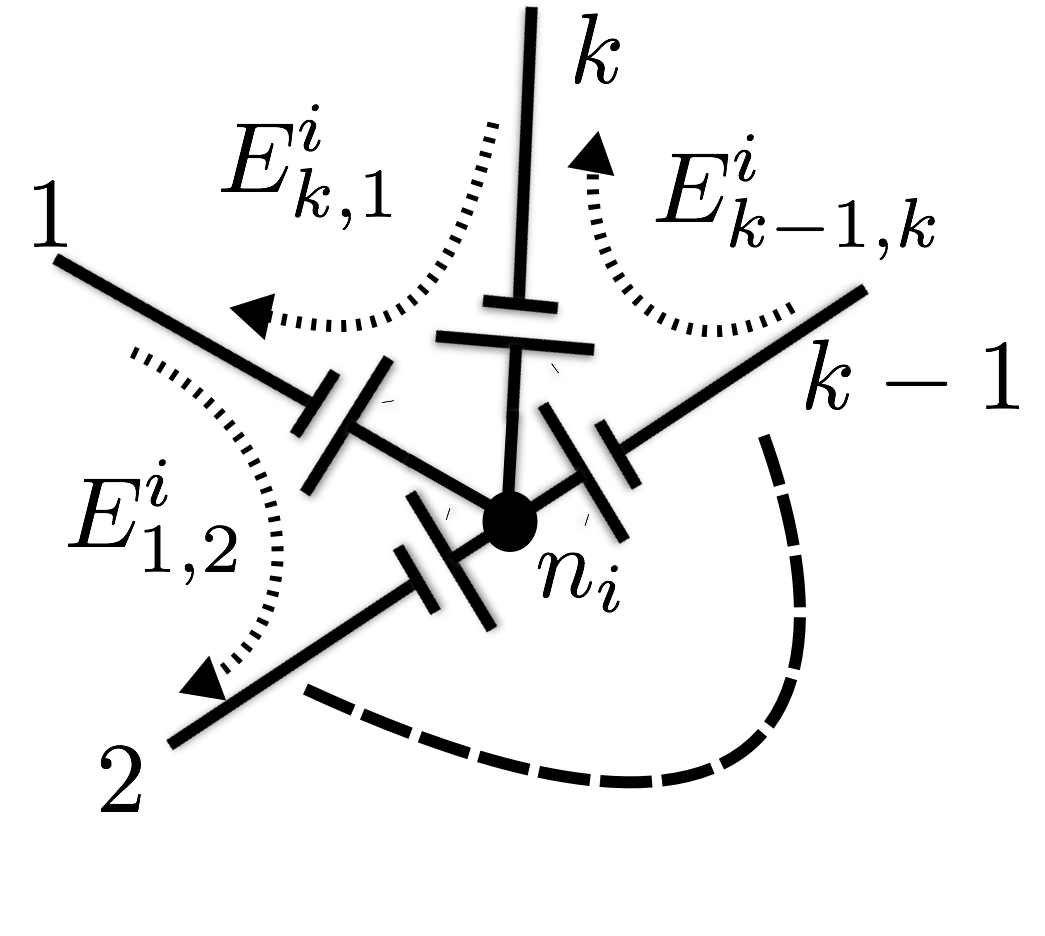}
\caption{The edge ordering attached to a node.}
\label{fig:order}
\end{figure}
For each edge $\beta$ (which represents a resistance) between the nodes $(n_i,n_j)$ and, and we consider a tuple of voltages $(\mathcal V_{\beta,i}, \mathcal V_{\beta,j})$ associated to it as in Fig. \ref{fig:edgeconf}. Let us call $F_i$ the number of resistances attached to the node $n_i$, which in graph theory is commonly called degree. 
We define also $ \mathcal V_{\beta,i}$
The goal of this section is to derive the voltages $\mathcal V_{\beta,z}$ based on the configuration of the spins, which as we will see is connected to the voltages $E^i_{\beta_1,\beta_2}$ below.  As introduced in \cite{gwc}, the node configuration can be assessed via the voltage integral across the node, starting from a resistance $\beta_1$ and going into a resistance $\beta_2$. In the formalism of the anisotropic magnetoresistance, given a certain local node $n_i$, and spin configuration at that node, a number of voltages $E^i_{\beta_1,\beta_2}$ can be obtained via the anisotropic magnetic effect:
\begin{equation}
V_{\gamma}=\int_{\gamma} \vec E\cdot d\vec t  %\nonumber \\
%=\int_{\gamma} \left(  \rho_0 \vec J+\hat m \Delta \rho (\hat m\cdot \vec J)\right)\cdot d\vec t\nonumber \\
= V_0+\Delta \rho \int_{\gamma}   \left(\hat m\cdot \vec J\right) \hat m\cdot d\vec t.
\label{eq:amr}
\end{equation}
Let us call $\mathcal G$ the graph that represents the circuit.
\subsubsection{Bulk}
If the graph is planar, then if $K$ is the number of resistances entering a node $i$, because of the planarity of the graph, only $K$ values of $E^i_{\beta_1,\beta_2}$ for a given node. This is due to the fact that for planar graphs only a number of cycles equal to the number of faces of the dual graph are necessary to obtain a self-consistent equation. However, the number of faces in this cases equals the number of entering edges. Thus, a very natural choice is to choose a set of fundamental loops in the circuit that are associated to each node in the dual graph. Also, because the graph is planar, we can choose a consistent orientation for each (fundamental) cycle in the circuit. 
Given this prescription for each node $n_i$, the number of integrands $E_{z,z+1}^i$ is equal to the number of voltages $\mathcal V_{\beta,i}$. In particular, we have the relationship, obtained by performing the integration via eqn. (\ref{eq:amr}), and the voltages $\mathcal V_{\beta,i}$. For each node $n_i$, let us call $\mathcal B_i$ the set of 
edges incident to that node. Because of the planarity, it is possible to give a consistent ordering to the edges $B_i=\{b_1^i,\cdots, b_{F_i}^i \}$ as well such that $b_{r+1}-b_r=1$, as in Fig. \ref{fig:order}. Then, we have
\begin{eqnarray}
\mathcal V_{b^i_{z},i}-\mathcal V_{b^i_{z+1},i}&=&E_{z,z+1}^i\ \ \forall\ 1 \leq z \leq F_i-1 \nonumber \\
\mathcal V_{b^i_{F_i},i}-\mathcal V_{b^i_{1},i}&=&E_{F_i,1}^i \ \ \ \ z=F_i\nonumber \\
\end{eqnarray}
where $F_i$ and $F_j$ are the number of resistances attached to the nodes $n_i$ and $n_j$ respectively.
It is not hard to see that the equation above, for each node, can be written in the more compact form:
\begin{equation}
\ _{F_i} D \vec V_{\cdot, i}= \vec E^i
\label{eq:sol}
\end{equation}
where the matrix $\ _{F} D$ is a matrix of size $F\times F$ given by:
\begin{equation}
\ _{F} D=\left(\begin{array}{ccccc}
1 & -1 & 0 & \cdots & 0 \\
0 & \ddots & \ddots & \ddots & \vdots \\
\vdots & \ddots& \ddots  & \ddots &  0\\
0 & \cdots& 0  & 1 &  -1\\
-1 & 0 & \cdots & 0 &  1
\end{array} \right),
\end{equation}
which is clear to be the discrete derivative on a circle with $F$ points. Thus, it is clear that for each node, this matrix is not invertible as it contains one null eigenvalue, with eigenvector proportional to $\vec e_i= \left(1,\cdots,1 \right)^t$ of arbitrary sizes $F_i$.  We are thus left with an ill-defined problem. The invertible subspace has dimension $F_{i}-1$, and we have thus the freedom of writing the solution of eqn. (\ref{eq:sol}) as:
\begin{equation}
{\vec {\mathcal V}}_{\cdot, i}=\ _{F_{i}} \tilde D^{- 1} \vec E^i + c_i \vec 1
\end{equation}
for an arbitrary constant $c_i$ associated with each node, and where we called $\tilde D^{-1}_i$ the pseudo-inverse operator. As we will see however the choice of this constant does not have any physical implication and we can freely set it to zero.
The pseudo-inverse for the forward difference operator can be written as $\ _F \tilde D^{-1}=(\ _F \tilde D^t \ _F \tilde D)^{-1}\ _FD$, where $\ _F \tilde D^{-2}$ is the pseudo-inverse of the second difference operator.
We focus for now on the pseudo-inverse $\ _{F_{i}} \tilde D^{- 1}$ of the matrix, that can however be explicitly calculated, as we know that $\ _{F} D \ _{F} D^t= \ _{F} D^t \ _{F} D=\ _{F} D^2$, e.g. the discrete second derivative on the circle of dimension $F$. %The non-zero eigenvalues of $\ _{F_i} D^2$ are $\ _i \lambda_k=- \frac{k^2 \pi^2}{F_i^2}$ if $k$ is even, and $\ _i\lambda_k=- \frac{(k-1)^2 \pi^2}{F_i^2}$ if $k$ is odd. The normalized non-null eigenvectors are $\ _iv^k_j=\frac{2}{F_i} \text{sin}\left( \frac{k \pi j}{F_i}\right)$ if $k$ is even and $\ _iv^k_j=\frac{2}{F_i} \text{cos}\left( \frac{ (k-1) \pi j}{F_i}\right)$ if $k$ is odd. 
 Thus, for each spin configuration of the spin ice at each node, given by the associated voltages $\vec E^i$, we have a vector of the effective voltages on each edge $\beta$, which can be written as
\begin{equation}
V_{\beta}=\mathcal V_{\beta,i}- \mathcal V_{\beta,j }+  q_\beta,
\end{equation}
where $q_\beta=c_i -c_j$. 
 \subsubsection{Fixing of boundary resistances}
 \begin{figure}
\includegraphics[scale=1]{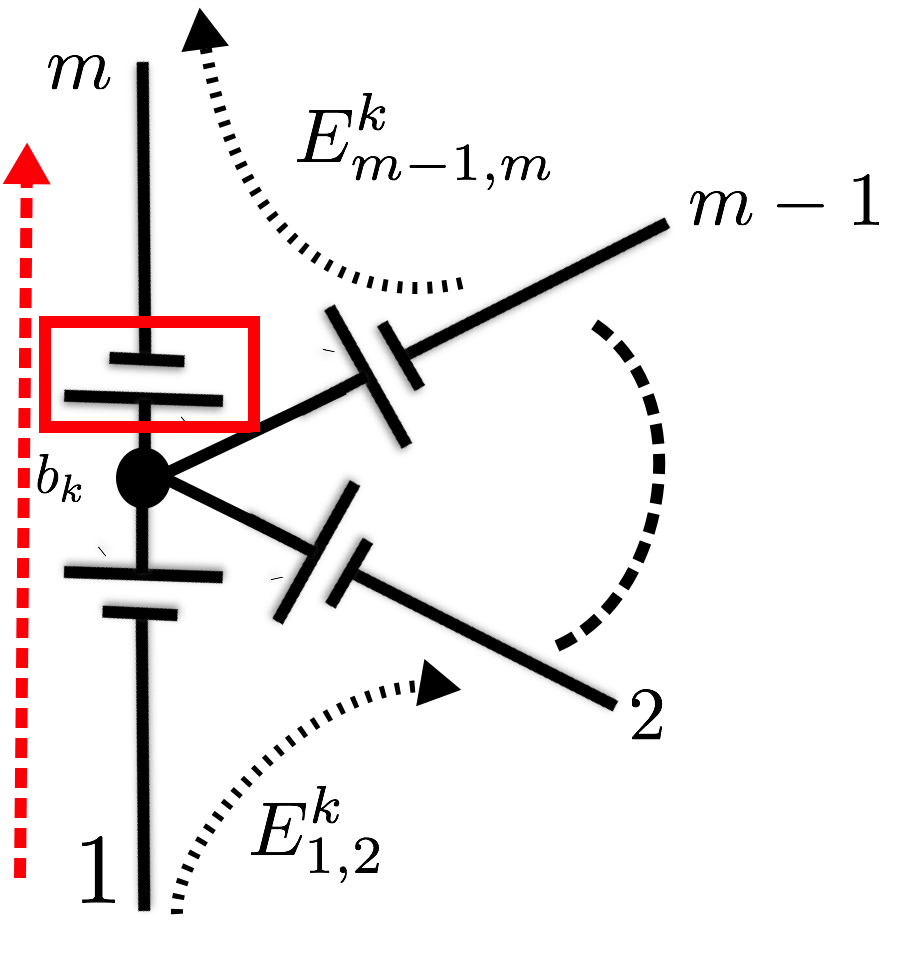}
\caption{Fixing of boundary resistances via setting to zero a voltage.}
\label{fig:boundary}
 \end{figure}
 The inversion problem at the boundary is slightly more complicated than the one in the bulk, and requires the prescription of setting some voltages to zero to avoid overdetermination. Given a circuit with a well-identifiable boundary, the it is not hard to see that given the prescription of Fig. \ref{fig:edgeconf}, the system of equations is underdetermined. At each node on the boundary with $m$ resistances, we have $m-1$ loop constraints. Thus, we need to find a way to reduce the number of voltages on the boundary for each node. Our prescription is the following. Let us consider the set of boundary resistances $\mathcal R_b=\partial \mathcal G$. Because the graph is assumed to be planar, we can assign a consistent orientation to this boundary, $\mathcal O$. Then, our prescription is that, given $\mathcal V_{\beta,k}$ if the orientation of the boundary and the positive side of the voltage generator agrees, then we keep it, while otherwise we remove it (or set it to zero). For instance, in Fig. \ref{fig:boundary}, given the orientation of the boundary (red arrow), the generator highlighted in red is set to zero. It is not too hard to see that this prescription removes the extra degree of freedom at each node on the boundary.

\subsection{C. General approach: absorbing the spin configurations in voltage sources}

As it is shown in the Appendix, given the node dependent voltage configurations of eqn. ({eq:amr}), we can obtain \textit{equivalent} voltage generators depending on the configurations of the spins $\vec V(\{s_i \})$. This is important, as we can now
write an exact equation for the currents of the system at equilibrium,  as this is a resistive system with voltages in series. The solution is known and given by:
\begin{equation}
\vec i=- A^t (A R A^t )^{-1} A \vec V,
\end{equation}
where $A$ is the directed cycle matrix on the fundamental cycles of the circuit. Here we assume that the voltage $\vec V$ is indeed depending on the internal spin configuration $\vec s$.

We now comment on the constants $c$'s. It is interesting to note that these can be written as $\vec q=B^t \vec c$, where $B^t$ is the directed incidence matrix of the graph. 
However, it is known that $A B^t=0$, and thus any configuration of these constants has no impact on the configuration of the currents, as one would expect from a change in potential. Another way to see this is by noticing that for each fundamental loop, necessarily at each node the same constant must be counted twice. However, since the cycle is directed, via the Kirchhoff law the same constant appears twice but with opposite signs, as it can be seen in Fig. \ref{fig:order}. We can thus set these to zero.

Let us now discuss how to write the effective memory of the component. The voltage $\vec V$ is $n+1$ dimensional, where $n$ is the number of edges internal to the device, and $1$ is the edge where the external voltage is applied.
First, let us call the matrix $Q=-A^t (A R A^t )^{-1} A$. The diagonal matrix $R$ can be written as $\text{diag}(r,\cdots,r,R_v)$, where $r$ is the resistance of a single alloy nanowire, while $R_v$ is the resistance of the battery.  For the matrix $Q$, we consider the following splitting:
\begin{equation}
    Q=\left(
    \begin{array}{cc}
         Q_{00} & \vec Q^t  \\
         \vec Q & Q_r 
    \end{array}
    \right),
    \label{eq:split}
\end{equation}
which is necessary to distinguish the resistance of the device, and the resistance of the battery.
Let us call $(\vec V)_0=v_0$ the applied voltage, and the rest $n$-dimensional vector $\vec V_r$, which are the internal voltages. Similarly, we introduce the splitting of the currents $\vec i$, as $(\vec i)_0=i_0$ and $\vec i_r$, the $n-$dimensional vector associated with the internal currents. Clearly, at equilibrium, we must have that these voltages depend linearly on the magnetic anisotropic effect, and on the internal configuration of the spins. We can write, because of eqn. (\ref{eq:amr}), 
\begin{equation}
    \vec V_r=\Delta \rho M(s) \vec i_r.
\label{eq:amrcurr}
\end{equation}
The equation above can be written, given the splitting of eqn. (\ref{eq:split}), as
\begin{eqnarray}
i_0&=&Q_{00} v_0+ \vec Q\cdot(\Delta \rho M(s) \vec i_r) \nonumber \\
\vec i_r&=&v_0 \vec Q+ \Delta \rho Q_r M(s) \vec i_r.
\end{eqnarray}
The internal currents at equilibrium are thus:
\begin{equation}
\vec i_r=v_0 \left(I-\Delta \rho Q_r M(s)\right)^{-1} \vec Q.
\end{equation}
Using the equation above for internal currents, we have
\begin{equation}
    i_0= Q_{00} v_0 + v_0 \vec Q\cdot\left(h M(s)\left(I-\Delta \rho Q_r M(s)\right)^{-1}\right)\vec Q.
\end{equation}
Thus, at the first order in $\Delta \rho$, we obtain that
\begin{equation}
    \frac{i_0}{v_0}=Q_{00}+ \Delta \rho \vec Q^t  M(s)\vec Q +O(h^2)
    \label{eq:theq}
\end{equation}
It is not hard that we can re-write eqn. (\ref{eq:theq}) in terms of resistances. We have
\begin{equation}
    \frac{1}{R}=\frac{1}{R_0}+\frac{1}{R_m(s)},
\end{equation}
where $R_0$ is the resistance when no anisotropic magnetic resistance is present. 
\begin{figure}
    \centering
    \includegraphics{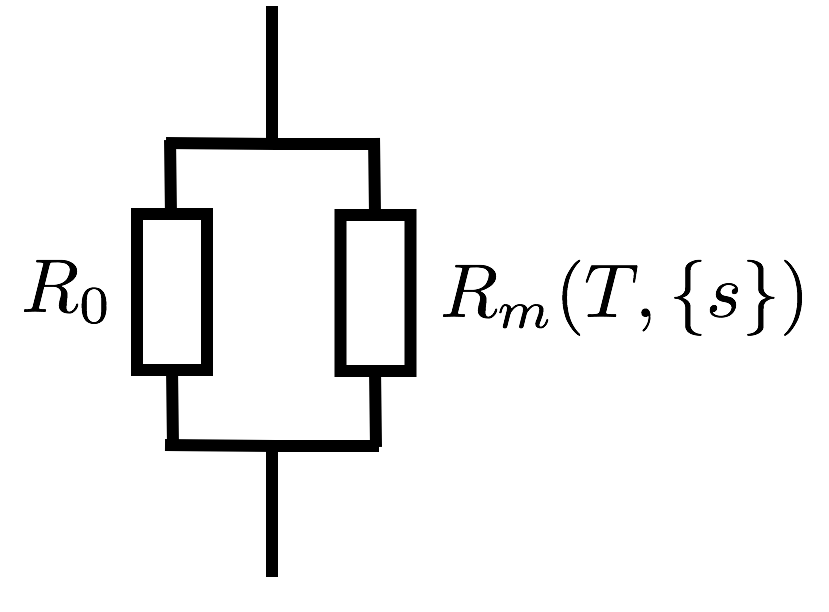}
    \caption{Effective resistance due to the anisotropic magnetoresistance effect, which induces a resistive state which depends on the internal spin state.}
    \label{fig:parallel}
\end{figure}
It is also interesting to note that the contribution to the conductance is a quadratic form.
We see that the formula above states that the effective conductances due to the magnetic anisotropy and conductance of the alloys at zero external field sum. We note that however $M(s)$ can change in time due to the currents. 
Thus, the equation above states how the effective resistance changes with the internal degrees of freedom. For $h\rightarrow 0$, the effective resistance due to the magnetic anisotropy goes to infinity, and since these are in parallel, the resistance of the material goes to its original value.

It is thus now the goal to construct the matrix $M(s)$.
 
\subsection{D. A worked out example: mapping of the spin configuration to the voltages of Kagome ice}
\label{sec:sd}
\begin{figure}[ht!]
    \centering
    \includegraphics[scale=0.5]{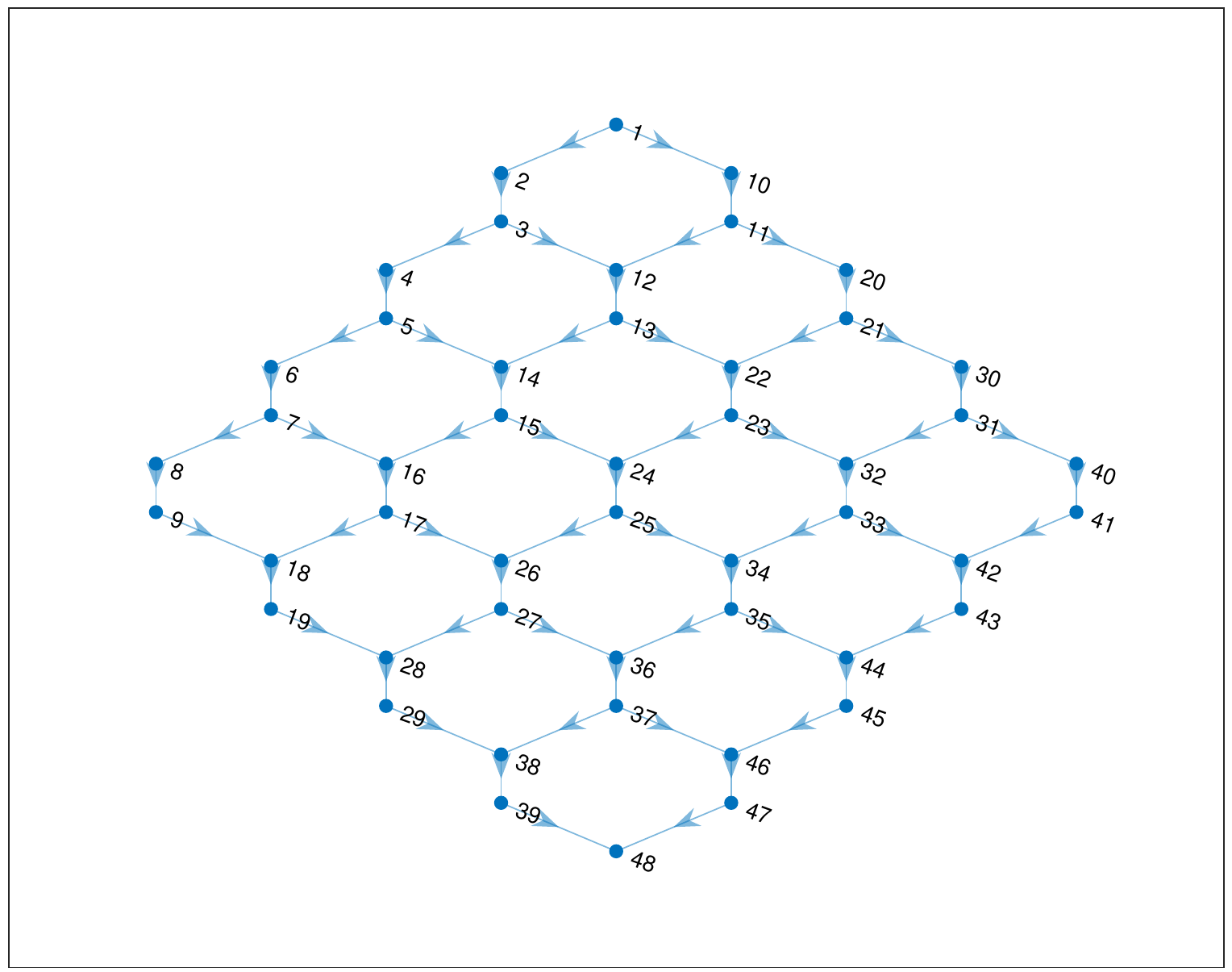}
    \caption{The Kagome lattice.}
    \label{fig:kagome}
\end{figure}

Let us consider the case of nodes with 3 legs, in the case of the Kagome lattice in Fig. \ref{fig:kagome}.

\begin{figure}[b!]
    \centering
    \includegraphics[scale=0.3]{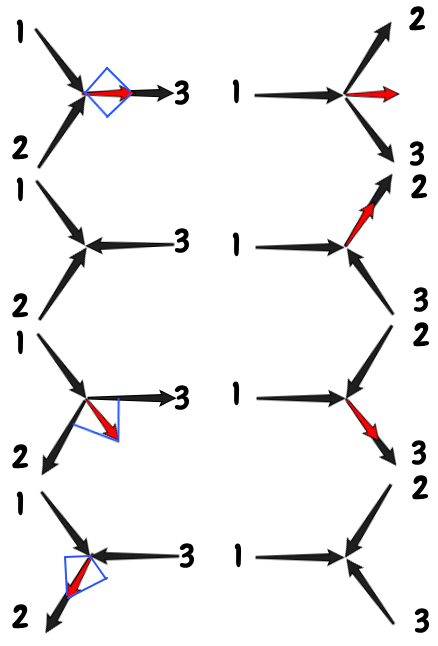}
    \caption{The 2-1 and 1-2 nodes, and the associated direction of the node magnetization for each associated spin configuration.}
    \label{fig:proj}
\end{figure}

As we have seen, this can be done node by node. For each node, we are going to have, given a configuration of the three spins incoming to that node, a voltage configuration which depends on the spins $s_1,s_2,s_3$. For clarity, $s_i$ is positive if it points towards the right  We know that the voltage configuration is independent from a change of the sign of the three spins, as this results in a change of direction of the magnetization at the node, and the voltage drop is independent with respect to $\hat m \rightarrow-\hat m$. For each edge, we are going to have four possible configurations of the spins at the node:
\begin{eqnarray}
   &(1)&\ \{s_1,s_2,s_3\},\{-s_1,-s_2,-s_3\}\nonumber \\
&(2)&\     \{s_1,s_2,-s_3\},\{-s_1,-s_2,s_3\}\nonumber \\
&(3)&\     \{s_1,-s_2,s_3\},\{-s_1,s_2,-s_3\}\nonumber \\
&(4)&\     \{s_1,-s_2,-s_3\},\{-s_1,s_2,s_3\}.\nonumber 
\end{eqnarray}
Thus, given a 3-dimensional vector $\vec E$ for a node for each of the four configurations above as:
\begin{eqnarray}
    \vec E(s_1,s_2,s_3)&=&\vec E^{(1)}\left(\delta_{s_1}\delta_{s_2}\delta_{s_3}+\delta_{s_1}\delta_{s_2}\delta_{s_3}\right) \nonumber \\
    &+&\vec E^{(2)}\left(\delta_{s_1}\delta_{s_2}\delta_{-s_3}+\delta_{-s_1}\delta_{-s_2}\delta_{s_3}\right) \nonumber \\
    &+&\vec E^{(3)}\left(\delta_{s_1}\delta_{-s_2}\delta_{s_3}+\delta_{-s_1}\delta_{s_2}\delta_{-s_3}\right) \nonumber \\
    &+&\vec E^{(4)}\left(\delta_{s_1}\delta_{-s_2}\delta_{-s_3}+\delta_{-s_1}\delta_{s_2}\delta_{s_3}\right)
\end{eqnarray}
where $\delta_{s}$ is a Kronecker delta which is one if $s=1$ and zero if $s=-1$. This Kronecker delta can be written $\delta_s=\frac{1-s}{2}.$
It is now not hard to see that we have:
\begin{eqnarray}
    \vec E_2(s_1,s_2,s_3)&=& \frac{\vec E^{(1)}+\vec E^{(2)}+\vec E^{(3)}+\vec E^{(4)}}{4} \nonumber \\
    &+& \frac{\vec E^{(1)}+\vec E^{(2)}-\vec E^{(3)}-\vec E^{(4)}}{4}s_1 s_2 \nonumber \\
    &+&\frac{\vec E^{(1)}-\vec E^{(2)}+\vec E^{(3)}-\vec E^{(4)}}{4}s_1 s_3 \nonumber \\
    &+&\frac{\vec E^{(1)}-\vec E^{(2)}-\vec E^{(3)}+\vec E^{(4)}}{4}s_2 s_3
\end{eqnarray}
On a Kagome lattice we have two type of nodes. Let us call them $1\rightarrow 2$ and $2\rightarrow 1$, as in Fig. \ref{fig:proj}. We assume that in both cases the currents direction are from the left to the right, that the integration over the cycles are clockwise and that the spins are positive if they point right.
We work first with the 2-1 node.
We call $s_1$ and $s_2$ the in-nodes and $s_3$ the out-node.

In this case,
\begin{eqnarray}
    \vec E^{(1)}_{2-1}&=&\left(\begin{array}{c}
    E_{12}= \frac{i_1-i_2}{2}\Delta\rho\\
    E_{13}= -\frac{2i_3+i_1}{2}\Delta\rho \\
    E_{23}= \frac{i_2+2i_3}{2}\Delta\rho
    \end{array}\right),\nonumber \\
    \vec E^{(2)}_{2-1}&=&\left(\begin{array}{c}
    E_{12}=0 \\
    E_{13}=0 \\
    E_{23}=0
    \end{array}\right),\nonumber \\
    \vec E^{(3)}_{2-1}&=&\left(\begin{array}{c}
    E_{12}= \frac{2i_1-i_2}{2}\Delta\rho \\
    E_{13}= -\frac{i_3+2i_1}{2}\Delta\rho \\
    E_{23}= \frac{-i_2+i_3}{2}\Delta\rho
    \end{array}\right),\nonumber \\
    \vec E^{(4)}_{2-1}&=&\left(\begin{array}{c}
    E_{12}=\frac{i_1-2i_2}{2}\Delta\rho\\
    E_{13}=-\frac{i_1+i_3}{2}\Delta\rho \\
    E_{23}=\frac{2i_2+i_3}{2}\Delta\rho \\
    \end{array}\right). 
\end{eqnarray}
In the case $1-2$ instead, we have
\begin{eqnarray}
    \vec E^{(1)}_{1-2}&=&\left(\begin{array}{c}
    E_{12}= -\frac{2i_1+i_2}{2}\Delta\rho\\
    E_{13}= \frac{2i_1+i_3}{2}\Delta\rho \\
    E_{23}= \frac{-i_3+i_2}{2}\Delta\rho
    \end{array}\right),\nonumber \\
    \vec E^{(2)}_{1-2}&=&\left(\begin{array}{c}
    E_{12}=-\frac{2i_2+i_1}{2}\Delta\rho \\
    E_{13}=\frac{i_1+i_3}{2}\Delta\rho \\
    E_{23}=\frac{-i_3+2i_2}{2}\Delta\rho
    \end{array}\right),\nonumber \\
    \vec E^{(3)}_{1-2}&=&\left(\begin{array}{c}
    E_{12}= \frac{i_1+i_2}{2}\Delta\rho \\
    E_{13}= \frac{2i_3+i_1}{2}\Delta\rho \\
    E_{23}= \frac{-2i_3+i_2}{2}\Delta\rho
    \end{array}\right),\nonumber \\
    \vec E^{(4)}_{1-2}&=&\left(\begin{array}{c}
    E_{12}=0\\
    E_{13}=0 \\
    E_{23}=0 \\
    \end{array}\right). \nonumber
\end{eqnarray}
The factors of $\frac{1}{2}$ come from the projection onto the current directions ($\cos(\frac{\pi}{3})=\frac{1}{2}$). The effective magnetic moment at the node is in fact either directed towards the link, or has an angle $\frac{\pi}{3}$. We stress that the signs of the currents depend only on the direction of the integration of the voltage over the node with respect to the direction of the currents, and not on the magnetization.

The voltage at each link, since it is the difference of two voltage sources, depends on five different spins, which decide the magnetization of the nearby nodes.

We have that 
\begin{eqnarray}
    V_\beta&=& \mathcal V_{\beta,i}-\mathcal V_{\beta,j} \nonumber \\
    &=&\tilde D_\beta^{-1}\big(
    \frac{\vec E^{(1)}_i+\vec E^{(2)}_i+\vec E^{(3)}_i+\vec E^{(4)}_i}{4} \nonumber \\
    &+& \frac{\vec E^{(1)}_i+\vec E^{(2)}_i-\vec E^{(3)}_i-\vec E^{(4)}_i}{4}s_1 s_2 \nonumber \\
    &+&\frac{\vec E^{(1)}_i-\vec E^{(2)}_i+\vec E^{(3)}_i-\vec E^{(4)}_i}{4}s_1 s_3 \nonumber \\
    &+&\frac{\vec E^{(1)}_i-\vec E^{(2)}_i-\vec E^{(3)}_i+\vec E^{(4)}_i}{4}s_2 s_3\big) \nonumber \\
       &-& D^{-1}_\beta\big(\frac{\vec E^{(1)}_j+\vec E^{(2)}_j+\vec E^{(3)}_j+\vec E^{(4)}_j}{4} \nonumber \\
    &+& \frac{\vec E^{(1)}_j+\vec E^{(2)}_j-\vec E^{(3)}_j-\vec E^{(4)}_j}{4}s_3 s_4 \nonumber \\
    &+&\frac{\vec E^{(1)}_j-\vec E^{(2)}_j+\vec E^{(3)}_j-\vec E^{(4)}_j}{4}s_3 s_5 \nonumber \\
    &+&\frac{\vec E^{(1)}_j-\vec E^{(2)}_j-\vec E^{(3)}_j+\vec E^{(4)}_j}{4}s_4 s_5 \big)  \Big).
\label{eq:kagomerules}
\end{eqnarray}
From the expression above we see that $M(s)$ is not-diagonal, but that for the Kagome lattice is a block which involves 5 currents.
%It is however interesting to note the following. In the ground state we have $D^{-1}(E^{(1)}_i-E^{(2)}_j)$\textcolor{red}{add text}

\begin{figure}
    \centering
    \includegraphics[scale=0.3]{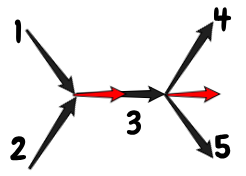}
    \caption{A certain configuration of a node.}
    \label{fig:node}
\end{figure}

One immediate example is an horizontal edge in the ground state. The non-zero portion of the matrix $M(s)$ which corresponds to the voltage $V_3$ and the currents $i_1,\cdots, i_5$ as in Fig. \ref{fig:node} is given by:

\begin{eqnarray}
M(s)=\begin{array}{c}
i_1 \\
i_2\\
i_3\\
i_4\\
i_5
\end{array}\left(
\begin{array}{c}
 -\frac{s_1 s_2}{12}+\frac{s_1 s_3}{12}+\frac{13}{24} \\
 -\frac{s_1 s_3}{12}+\frac{s_2 s_3}{6}+\frac{5}{24} \\
 \frac{s_1 s_3}{6}+\frac{s_2 s_3}{6}+\frac{s_3 s_4}{12}+\frac{s_3
   s_5}{12}+\frac{s_4 s_5}{6}+\frac{5}{6} \\
 \frac{s_3 s_4}{6}+\frac{1}{6} \\
 \frac{s_3 s_4}{12}-\frac{s_3 s_5}{6}-\frac{1}{12} \\
\end{array}
\right)
\end{eqnarray}
We see from the expressions above that this formalism is ought to be used for a numerical simulation rather than for analytical computations, and that $M(s)$ depends on pairs of variables $s_i s_j$.

\subsubsection{Magneto-resistance memory}
We now consider the simplest non-trivial example of magnetoresistance memristor device.
In this simple example we focus on a simple enough case for which most of the techniques we have developed for general constructions are not necessary. In particular, only one voltage per edge is necessary, and thus simply the extra voltage vector can be simply written as
\begin{equation}
    \vec V_{mrs}=\Delta\rho \tilde D^{-1} \vec E_{2,1}(s_1,s_2,s_3)=\Delta\rho \tilde D^{-1} M(s) \vec i
\end{equation}
and if we use the equilibrium current equation:
\begin{equation}
    \vec i=-Q (\vec V_0+\vec V_{mrs})=- Q \vec V_0-\delta \rho Q D^{-1} M(s) \vec i:
\end{equation}
from which we obtain
\begin{eqnarray}
    \vec i&=& -(1-\Delta\rho Q \tilde D^{-1} M(s))^{-1}Q \vec V_0 \nonumber \\
    &\approx&-\left(1+\Delta\rho Q \tilde D^{-1} M(s)\right)Q \vec V_0
\end{eqnarray}
Upon investigation, we find that the product of the matrices $M(s)$ and $\tilde D^{-1}$ are given by:
%\begin{widetext}
\begin{equation}
D^{-1}M(s)=\left(
\begin{array}{ccc}
 \frac{1}{24} \left(2 s_1 \left(s_3-s_2\right)+13\right) & \frac{1}{12} \left(-3 s_2 s_3+s_1
   \left(s_2+s_3\right)-9\right) & \frac{1}{12} \left(-\left(s_1+s_2\right) s_3-5\right) \\
 \frac{1}{12} \left(2 s_1 \left(s_2-s_3\right)-13\right) & \frac{1}{24} \left(2 s_2 \left(s_3-s_1\right)+13\right) &
   \frac{1}{12} \left(-\left(s_1+s_2\right) s_3-5\right) \\
 \frac{1}{24} \left(2 s_1 \left(s_3-s_2\right)+13\right) & \frac{1}{24} \left(5-2 \left(s_1-2 s_2\right) s_3\right)
   & \frac{1}{6} \left(\left(s_1+s_2\right) s_3+5\right) \\
\end{array}
\right)
\end{equation}
%\end{widetext}
which we will now use.
\begin{comment}
\begin{eqnarray}
D^{-1}&=&\left(
\begin{array}{ccc}
 \frac{1}{3} & 0 & -\frac{1}{3} \\
 -\frac{1}{3} & \frac{1}{3} & 0 \\
 0 & -\frac{1}{3} & \frac{1}{3} \\
\end{array}
\right) \nonumber \\
M(s)&=&\left(
\begin{array}{ccc}
 \frac{1}{8} \left(2 s_1 \left(s_3-s_2\right)+13\right) & \frac{1}{8} \left(2 s_2(s_1 - s_3)-13\right) & 0 \\
 \frac{1}{8} \left(2 s_1 \left(s_2-s_3\right)-13\right) & 0 & -\frac{1}{4} \left(s_1+s_2\right) s_3-\frac{5}{4} \\
 0 & \frac{1}{8} \left(s_3( 4 s_2-2 s_1) +5\right) & \frac{1}{4} \left(\left(s_1+s_2\right) s_3+5\right) \\
\end{array}
\right)
\end{eqnarray}
\end{comment}
We see that the matrix which couples the internal spins to the internal currents is a rather non-trivial matrix which, however, depends only on the internal configuration. In the next section we show that when the spins are allowed to flip thermally, a non-trivial memory effect ar ises out of equilibrium.

Thus, the effective voltage in this case is simply
\begin{equation}
    \vec V(s)=\vec v_0+D^{-1} M(s) \vec i
\end{equation}
As a result, we have
\begin{equation}
    \vec V(s)=\left(
\begin{array}{c}
 \left(\frac{1}{12} i_3 ((-s_1-s_2) s_3-5)+\frac{1}{24}
   i_1 (2 s_1 (s_3-s_2))+13)+\frac{1}{12} i_2 (-3 s_2
   s_3+s_1 (s_2+s_3)-9)\right) \Delta \rho  \\
 \left(\frac{1}{12} i_3 ((-s_1-s_2) s_3-5)+\frac{1}{24} i_2
   (2 s_2 (s_3-s_1)+13)+\frac{1}{12} i_1 (2 s_1
   (s_3-s_2)+13)\right) \Delta \rho  \\
 v_0+\left(\frac{1}{24} i_2 (5-2 (s_1-2 s_2) s_3)+\frac{1}{6}
   i_3 ((-s_1-s_2) s_3-5)+\frac{1}{24} i_1 (2 s_1
   (s_3-s_2)+13)\right) \Delta \rho  \\
\end{array}
\right),
\end{equation}
which is the state-dependent effective voltage.

\subsection{E. Thermally induced flips: out of equilibrium properties}
It is interesting at this point to observe the out-of-equilibrium dynamics of the system.
We perform numerical simulations, apt at enhancing the effect and to show how a hysteresis loop typical of a memristive system emerges in this scenario.

At equilibrium, the currents satisfy the Kirchoff laws. For $\Delta \rho=0$, the system does not 
present any difference from a normal resistance. However, as $\Delta \rho\neq 0$, thermal coupling can affect the internal properties of the resistance.

Here we assume a very simple internal dynamics, governed by the thermal coupling due to the Joule heating of the device. The model we suggest is rather simple but explicative of the phenomenology.
The internal state of the device is assumed to evolve according to a Metropolis dynamics for the 3 spins, with a flipping probability:
\begin{equation}
    P(flip)\propto e^{-\frac{\Delta H}{T(t)}} 
\end{equation}
where $\Delta H$ is the energy difference between one configuration and the proposed one.
The energy is the simplest possible ferromagnetic coupling for 3 nanoislands, given by
\begin{equation}
    H=J\left( s_1 s_2+s_2s_3+s_1 s_3\right).
\end{equation}

The coupling between the internal states and the currents occurs via Joule heating:
as the currents flows, we assume a temperature for the devices which follows a very simple relationship:
\begin{equation}
    \frac{ d T(t)}{dt}=\frac{R (i_1^2+i_2^2+i_3^2)}{C_v m}- \sigma T^4,
\end{equation}
where the first term is due to the Joule heating effect (and we assume that the temperature is just the average heating of the three branches), and as a balancing effect for the temperature we consider Stefan's radiation law.

\begin{figure}
    \centering
    \includegraphics[scale=0.45]{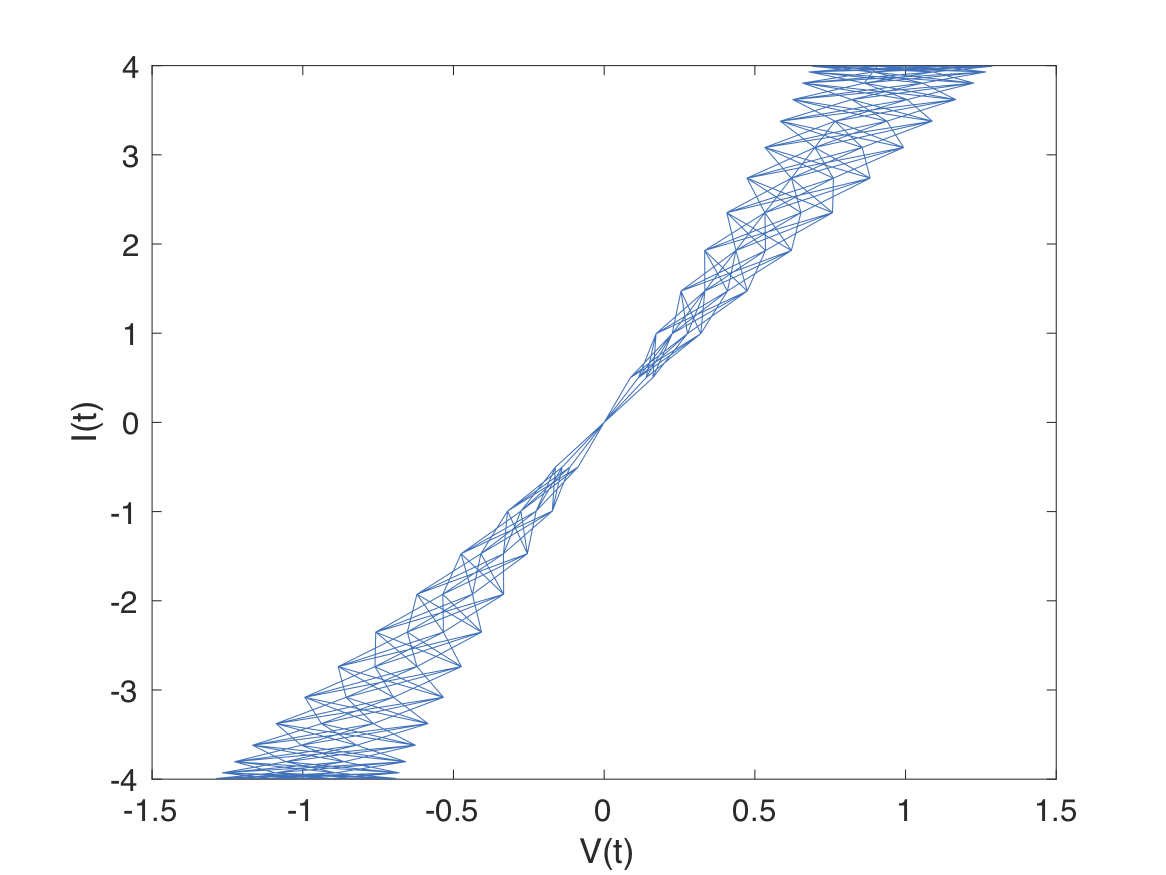}\\
    \caption{The hysteretic jumps which arises from the internal spin dynamics in the simple model of Fig. \ref{fig:simple}.}
    \label{fig:sim}
\end{figure}

Given this simple mechanism, we consider the out-of-equilibrium voltage $v(t)=v_0 \sin(\omega t)$, which is shown in Fig. Fig. \ref{fig:sim}. We observe zero-crossing hysteretic jumps due to the state of the spins, between two resistance lines. In order to observe a real memristive behaviour, we need to go to larger systems.

% I REMOVED the average because I didn't say much to me: the system is probably just too small.
%The result of the simulations are shown in Fig. \ref{fig:sim}. We plot the behavior of a single Monte Carlo versus the average over 1000 samples with identical parameters. While for a single simulation we see a behavior which is hysteretic but unclear. However, the average over all possible simulations show no clear hysteresis is present.

\subsubsection{Effective model after thermal averaging}
For a larger lattice, obtaining the matrix $M(s)$ can be challenging. However, the key features of the resistance can be inferred from thermal averaging as follows. Let us consider eqn. (\ref{eq:theq}) again:
$$    \frac{i_0}{v_0}=Q_{00}+ \Delta \rho \vec Q^t  M(s)\vec Q. $$
In particular, we are interested in the thermal average of the equation above, e.g.
\begin{equation}
    \langle \frac{i_0}{v_0} \rangle_T=Q_{00}+ \Delta \rho \vec Q^t  \langle M(s) \rangle_T\vec Q, 
\end{equation}
where $\langle \cdot \rangle_T$ is the thermal average, over all the possible configurations of the system at temperature $T$. We note that 
\begin{equation}
    \langle M(s) \rangle_T
\end{equation}
is a matrix which, for a local Hamiltonian, is composed of products of neighboring spins only, of the form
\begin{equation}
    \langle s_i s_j \rangle_T
\end{equation}
where the distance $d(s_i,s_j)$ is of order one. Depending on the system of interest, this average will lead to different results depending on the geometric arrangements of the nanoislands. In particular, for lattices which exhibit with a sharp transition from a disordered to an ordered phase at low temperature, we can approximate
\begin{equation}
     \langle s_i s_j \rangle_T\approx \begin{cases}
     0 & T\gtrapprox T_c, \\
     1 & T\lessapprox  T_c.
     \end{cases}
\end{equation}
Thus, because the matrix $M(s)$ is composed only of products of pairs of neighboring spins, we can write
\begin{equation}
    \langle M(s) \rangle_T =\begin{cases}
         M_{>} & T>T_c \\
     M_{<} & T< T_c.
    \end{cases}
\end{equation}
and we can think of an effective interpolation between two limiting values of the resistance.
Given this feature, can write an approximate (first order contribution) equation for the thermal average of the effective resistance of the form:
\begin{equation}
    \langle R^{-1}\rangle_T=R_0^{-1}+\Delta \rho \left( \theta_k(T-T_c)R_{<}^{-2}+\theta_k(T_c-T)R_{>}^{-2}   \right)
\end{equation}
where $R_{<}$ and $R_{>}$ are two resistances which depend on the value of $M(s)$ and the geometry of the spin ice (which is contained in $\vec Q$), and are defined by
\begin{eqnarray}
    R_{<}&=&\vec Q\cdot M_{<}\vec Q \\
    R_{>}&=&\vec Q\cdot M_{>}\vec Q.
\end{eqnarray}
The function $\theta_k(x)$ is a smoothed Heaviside-theta function.
Thus, in the simplest possible approximation, we see that under the application of a small magnetic field, and of joint permalloy islands (nanowires), the resistance of the material can be assumed to be a phase-change type of material (or switches), in which the system has two resistance phases depending on the current, which as a matter of fact controls the temperature of the permalloy via Joule heating. If the transition is not sharp but only a crossover, then the it is not too daring to assume that
\begin{equation}
    \langle R^{-1}\rangle_T=R_0^{-1}+\frac{\Delta \rho}{  \tilde R(T)^{2}}   
\end{equation}
where $\tilde R(T=0)=R_{<}$ and $\tilde R(T=\infty)=R_{>}$ is a smooth function. In this case, it seems reasonable to assume that the material will fall in the thermal memristor framework introduced in \cite{thermal1}.

In this case, because we expect the typical memristive $V-I$ hysteresis to be small, it is possible to see the change in the resistances from the $v-r$ Lissajous figures, obtained from the functional dependence of the effective resistance which we have obtained.
\begin{figure}
    \centering
    \includegraphics[scale=0.37]{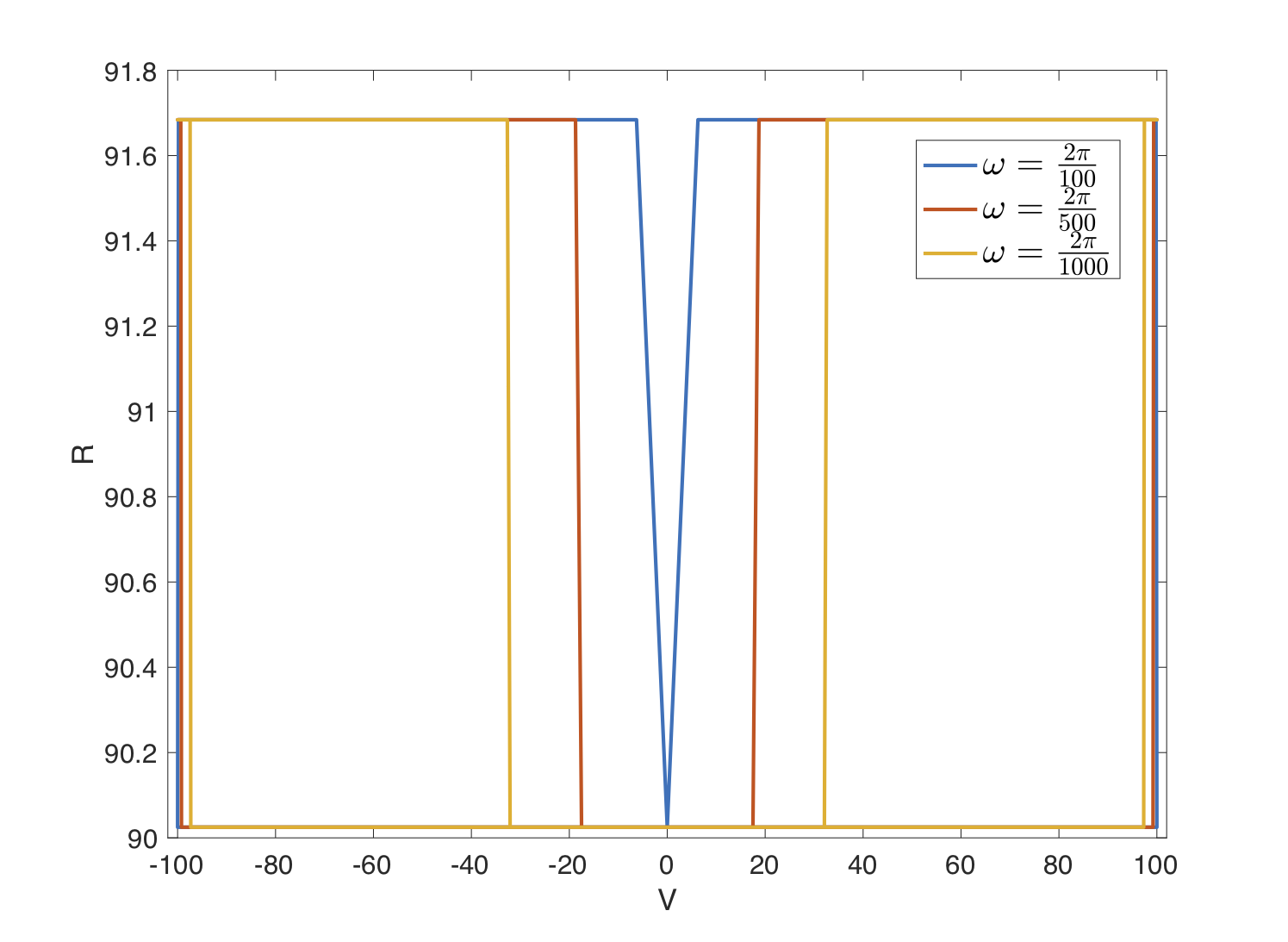}
    \caption{Lissajous figure for the resistance as a function of the voltage for a sharp ordering transition ($k=\infty$), for a sinusoidal input and as a function of the frequency of functional form $V=V_0 \sin(\omega t)$.}
    \label{fig:lissajouss}
\end{figure}
Albeit the exact numbers will depend on the type of material, we expect to be able to distinguish the type of transition from the $V-R$ curves of the device as a function of the frequencies. At slower frequencies, the changes in the resistance will be more symmetric in the continuous case, while more abrupt but still hysterestic in the discontinous case.